\documentstyle[12pt,epsf]{article}

\newcommand{\be}{\begin{equation}}
\newcommand{\ee}{\end{equation}}
\newcommand{\bea}{\begin{eqnarray}}
\newcommand{\eea}{\end{eqnarray}}
\newcommand{\ep}{\varepsilon}

\newcommand{\Li}[2]{{\mbox{Li}}_{#1}\left(#2\right)}
\newcommand{\sfrac}[2]{\frac{{}_{#1}}{{}^{#2}}}
\newcommand{\zet}[1]{\zeta_{#1}}

\begin{document}
% -----  \input{title}
\thispagestyle{empty}
 \begin{flushright}
 INLO--PUB--15/94\\
 DTP--94/86\\
 hep-ph/9410232\\[9mm]
 September 1994
 \end{flushright}
\vspace{30 mm}
\begin{center}
 {\bf \large
 Zero-threshold expansion of two-loop self-energy diagrams}
  \footnote{This research is partly supported by the EU under contract
   numbers CHRX-CT92-0004 and\\
   $\hspace*{7mm}$INTAS-93-744.}
\vspace{10 mm} \\
  F.A.~Berends$^{a,}$\footnote{E-mail address:
  berends@rulgm0.leidenuniv.nl},
  A.I.~Davydychev$^{b,}$\footnote{On leave from
  Institute for Nuclear Physics,
  Moscow State University, 119899 Moscow, Russia.\\
  $\hspace*{7mm}$E-mail address: davyd@vsfys1.fi.uib.no},
  V.A.~Smirnov$^{c,}$\footnote{E-mail address:
  smirnov@compnet.msu.su}
  and J.B.~Tausk$^{d,}$\footnote{Address
  after October 1st, 1994:
  Institut f\"ur Physik,
  Johannes Gutenberg Universit\"at,\\
  $\hspace*{7mm}$Staudinger Weg 7, D-55099 Mainz, Germany}
\vspace{10 mm} \\
$^{a}${\em
  Instituut-Lorentz,
  University of Leiden, \\
  P.O.B. 9506, 2300 RA Leiden, The Netherlands}
\vspace{3 mm} \\
$^{b}${\em
  Department of Physics, University of Bergen,\\
  All\'{e}gaten 55, N-5007 Bergen, Norway}
\vspace{3 mm} \\
$^{c}${\em
  Institute for Nuclear Physics,
  Moscow State University, \\
  119899 Moscow, Russia}
\vspace{3 mm} \\
$^{d}${\em
  Department of Physics,
  University of Durham,\\
  Durham DH1 3LE, England}
\end{center}
\vspace{12 mm}
\begin{abstract}
An algorithm is constructed to derive a small momentum
expansion for two-loop two-point diagrams in all cases
where, due to the presence of physical thresholds,
there are singularities at zero external momentum.
The coefficients of this ``zero-threshold'' expansion
are calculated analytically for arbitrary masses.
Numerical examples, using diagrams occurring in the
Standard Model, illustrate the convergence of the expansion
below the first non-zero threshold.
\end{abstract}
\newpage
\setcounter{page}{2}
\setcounter{footnote}{0}

% ---- \input{intro}
\begin{center}
{\bf 1. Introduction}
\end{center}

The increasing precision of experiments \cite{lefr,schaile} testing
the Standard Model (SM) of electroweak interactions will demand a
matching precision from the theoretical predictions. The present
LEP/SLC experiments measure the $Z$ mass, its total and partial
widths, and various asymmetries with high accuracy. Theoretical
predictions for these quantities depend on the unknown top quark and
Higgs boson masses so that some freedom is left to fit theory to
experiment. When knowledge of the top mass from direct detection
becomes available the possibility to adjust $m_t$ in the theoretical
predictions will be narrowed and the tests will become more stringent.
At present there are hints of a value for the top mass of
$174\pm10^{+13}_{-12}$~GeV \cite{CDF}.
% First error statistical, second systematic.

A similar situation holds for the $W$ mass. At present its measured
and predicted (from the Fermi constant $G_F$) values are in agreement
when suitable $m_t$ and $M_H$ values are chosen. However the expected
higher accuracy on $M_W$ from LEP2 measurements \cite{kats} will test
the theory more severely.

When at some point in the future discrepancies arise between
theoretical predictions and experiment it will be tempting to invoke
extensions of the SM. This can only be done convincingly when SM
predictions really fail. This brings us to the question how accurate
the present theoretical calculations are.

The one-loop predictions of the SM have been studied by many people
and are incorporated in various fitting programs in use at LEP1. Some
special second order effects are often included as well. Although in
principle this should be well under control, in practice there are
subtle differences in implementing the theory, which may cause small
deviations in theoretical predictions. An analysis of this situation
is at the moment in progress \cite{pas}.

For some quantities one-loop calculations are definitely not adequate
and thus the dominant parts of some second order contributions should
be added.
The clearest examples are various QED radiative corrections to the
$Z$-line shape \cite{ber} and to small angle Bhabha scattering
\cite{been,jadach,remiddi}. The former are known to the desired accuracy,
the latter not. For small angle Bhabha scattering one could for
instance be interested in double box diagrams \cite{osland}, which
involve massless and massive particles.
Examples of second order corrections in the non-QED part of the
theory are contributions to the $\rho$ parameter \cite{Bij,rho}
and to $Z\rightarrow b \bar{b}$ decay, for a review see \cite{kniehl}.
Here one encounters two-loop diagrams with many massive particles.
The present calculations consider some limiting situations of heavy
top mass or heavy Higgs mass, which allow the neglect of $M_Z$ or
$M_W$ masses.

In the future these types of approximations will become less
acceptable. Therefore it seems unavoidable that physics issues will
dictate the study of exact two-loop corrections to the electroweak
theory.

This is why during the last few years a number of studies have been
made into the question of massive two-loop diagrams, the purely
massless case being much better understood. It is a sensible strategy
to study firstly the simplest two-loop diagrams, the self-energy
diagrams. Not surprisingly, most papers deal with those diagrams.
{}From a physical point of view this can also be justified, since at the
one-loop level these diagrams are usually the most relevant ones.
Since it has been shown that all two-loop self-energy diagrams can be
reduced to scalar diagrams \cite{Weigl} (see also \cite{Krei-tens}),
it is sufficient to consider only the latter ones.

The general case of massive two-loop diagrams occurs to be very
complicated, even for two-point (self-energy) functions.
Exact results in terms of known special functions are known
for a restricted number of special cases only (some of them can
be found in \cite{Broadh,BFT,ST}). In all these examples at least one
of the masses is equal to zero. A difficult problem,
unknown at the one-loop level, arises
when a diagram
contains a three-particle cut where all particles are massive.
There are some arguments that this result expanded around
four dimensions is for general $k^2$ not expressible in
terms of known functions like polylogarithms \cite{sch}
(see also in \cite{Broadh} where a special case of such a
diagram was investigated).

To be able to deal with diagrams that have not been
evaluated exactly, a number of approaches have been developed,
and some useful results have been obtained.
One method is a reduction to a two-dimensional integral
representation, both for convergent \cite{Kreimer} and divergent
\cite{Berends}
cases\footnote{Analogous integral representations can be
written also for three-point two-loop diagrams \cite{Kreimer3}.}.
A number of one-dimensional integral
representations with more complicated integrands are available
\cite{Buza}.
Some other methods \cite{Japan,Freiburg} can also be employed,
although they involve representations with a larger number of
parametric integrals. In all these cases,
the final result is obtained by numerical integration.

When one is exclusively interested in numbers and less in the
analytic structure the above method is sufficient, although
there is always the question whether the numerical program works
correctly in all cases.
When one wants to
understand the analytic structure and also wants to have another,
independent method, series expansions in the variables of the
problem, i.e. $k^2$ and the masses, provide a good alternative.
Moreover, this approach can be very useful when some of the
masses are unknown, because it gives analytic expressions
for the coefficients of the expansion.

For a general massive scalar two-loop self-energy diagram the
following expansions in $k^2$ have been studied. For small $k^2$ an
algorithm was developed for the construction of the Taylor series
coefficients \cite{DT} (the three-point case was considered in
\cite{FT}\footnote{In ref.~\cite{FT}, conformal mapping and
Pad\'e approximations were also employed for numerical
study of the behaviour above the threshold.}).
The expansion is valid below the smallest
physical threshold. For large $k^2$ again an algorithm for the evaluation of
the $1/k^2$ expansion coefficients was developed \cite{DST}. This
is based on general results on asymptotic expansions of Feynman diagrams
\cite{as-ex} where the coefficients are conveniently expressed in terms
of more simple diagrams of a certain type.
This expansion is valid above the largest physical threshold.
Although general explicit expressions for coefficients of arbitrary order
have not been obtained in such a way,
for some special diagrams multiple series in $k^2$ and the
masses, with closed expressions for the coefficients, are
now available \cite{Buza}.
These multiple series expansions converge in certain regions
and correspond to generalized hypergeometric functions.

{}From this discussion it is clear that there are $k^2$ regions,
where no expansion is available. Namely, the behaviour
above the smallest but below the largest physical threshold
needs to be described.
It is the purpose of the present paper to start to fill this gap.
Clearly filling the most general gap would be the best.
But we are going to start by filling a smaller gap,
corresponding to the cases when the lowest physical threshold
is zero, so besides massive particles some massless particles
are present in the diagram.
These are cases when an ordinary Taylor expansion \cite{DT}
does not work because of infrared singularities connected
with the limit $k^2 \to 0$. By doing this, we cover all remaining
holes in the small momentum expansion.
Moreover, we expect that this specific case may be helpful
for the most general problem of describing threshold behaviour.

In this paper, we are going to examine the region between the
$k^2 = 0$ threshold and the first non-zero threshold.
It will be shown that general
results on asymptotic expansions of Feynman diagrams \cite{as-ex} can
also be employed in this case.

The actual outline of the paper is as follows. In section 2 the
algorithm for the expansion is constructed from the general theorem.
All possible zero-threshold cases are considered. The actual
calculation of several coefficients is presented in section 3.
Also a comparison with some known analytical results is made. Section
4 contains numerical comparisons for some complicated, analytically
unknown cases. Conclusions are given in section 5. In the appendices
we present some formulae for two-loop vacuum integrals and discuss
the evaluation of massless integrals with numerators.

\vspace{3mm}

% ---- \input{sec2}
\begin{center}
{\bf 2. Constructing the expansion}
\end{center}

\begin{figure}[b]
      \epsffile{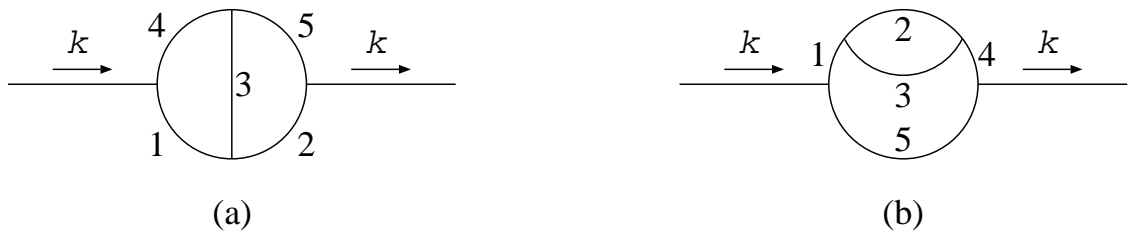}
      \caption[]{ Two-loop self-energy diagrams.}
\label{fig:diagrams}
\end{figure}

As in the previous papers \cite{DT,DST}, we shall consider here
two-loop Feynman integrals corresponding to the self-energy diagram
of Fig.~1a, denoting them as
\be
\label{defJ}
J\left( \{\nu_i\} ; \{m_i\} ; k \right)
= \int \int
  \frac{\mbox{d}^n p \; \mbox{d}^n q}
       {D_1^{\nu_1} D_2^{\nu_2} D_3^{\nu_3} D_4^{\nu_4} D_5^{\nu_5}} ,
\ee
where $n= 4-2\ep$ is the space-time dimension (in the framework
of dimensional regularization \cite{dimreg}), and $\nu_i$ are the powers
of the denominators $D_i \equiv p_i^2 - m_i^2 + i0$, $p_i$ being
the momentum of the corresponding line ($p_i$ are constructed from
the loop momenta $p$ and $q$ and the external momentum $k$).
We are mainly interested in all $\nu$'s being integer.
The cases when some of the $\nu$'s are zero correspond to reducing
lines in Fig.~1a to points. In such a way,
self-energy diagrams with four or three internal lines can
also be described. Moreover, by trivial decomposition (partial
fractioning) of the first and the fourth denominators (provided
that $\nu_1$ and $\nu_4$ are integer) we can reduce the integral
corresponding to Fig.~1b to the integrals (\ref{defJ}) with
$\nu_1$ or $\nu_4$ equal to zero (such a decomposition is required
only if $m_1 \neq m_4$). So, in the general case of self-energy
diagrams (with integer $\nu$'s) it is sufficient to consider only
the integrals (\ref{defJ}).

In general, the diagram in Fig.~1a has four physical thresholds.
Two of them correspond to two-particle cuts,
\be
\label{2pt}
k^2 = (m_1 + m_4)^2 \hspace{1cm} \mbox{and} \hspace{1cm}
k^2 = (m_2 + m_5)^2 ,
\ee
while the other two are connected with three-particle cuts,
\be
\label{3pt}
k^2 = (m_1 + m_3 + m_5)^2 \hspace{1cm} \mbox{and} \hspace{1cm}
k^2 = (m_2 + m_3 + m_4)^2 .
\ee
In special cases, some of these threshold values can be equal.

If all the thresholds (\ref{2pt})--(\ref{3pt}) are not equal to zero,
the integral (\ref{defJ}) is an analytic function in a vicinity of
$k^2 = 0$, so that the asymptotic expansion is
given by its Taylor series in $k^2$.
It is convergent when $k^2$ is less than the
value of the first physical threshold.
The general case of such an expansion of two-point functions
was considered in \cite{DT}, and some results for three-point functions
can be found in \cite{FT}. The situation is different, however, when
some of the physical thresholds (\ref{2pt})--(\ref{3pt})
become equal to zero. In this case, there can be
non-polynomial terms in $k^2$
as $k^2 \to 0$. Below we shall see that they appear as
$\ln(-k^2)$ terms as $n \to 4$. These are the ``zero-threshold''
cases that will be considered in the present paper.

For the diagram in Fig.~1a, four independent zero-threshold
configurations exist, namely
\footnote{2PT and 3PT mean two- and
three-particle thresholds, respectively.}:

Case 1: one zero-2PT (e.g. $m_2 = m_5 = 0$);

Case 2: two zero-2PT's ($m_1 = m_2 = m_4 = m_5 = 0$);

Case 3: one zero-3PT (e.g. $m_2 = m_3 =m_4 = 0$);

Case 4: one zero-2PT and one zero-3PT (e.g. $m_2 = m_3 = m_4 = m_5 = 0$).

\begin{figure}[htb]
 \[
 \begin{array}{cccc}
 \epsffile{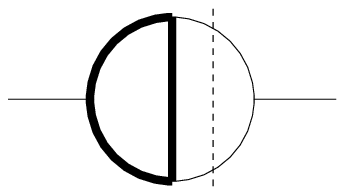} &
 \epsffile{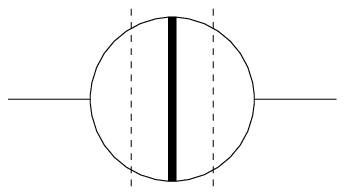} &
 \epsffile{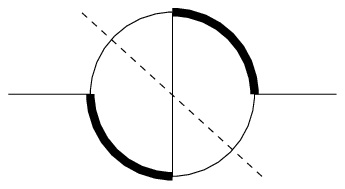} &
 \epsffile{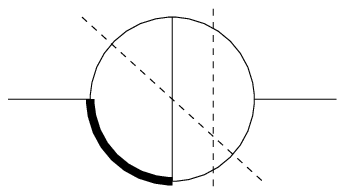} \\
 \mbox{Case 1} &
 \mbox{Case 2} &
 \mbox{Case 3} &
 \mbox{Case 4}
  \end{array}
  \]
      \caption[]{ Different zero-threshold configurations for the
         diagram in Fig.~1a.}
\label{fig:cuts}
\end{figure}

These cases are shown in Fig.~\ref{fig:cuts} where bold lines are massive,
narrow lines are massless, and the cuts corresponding to
zero-thresholds are denoted by dashed lines.
Note that, in case 1, having one more vanishing mass (e.g. $m_3$
 or $m_4$) does not produce a new zero-threshold configuration,
and the corresponding cases can be considered together with case 1,
e.g.:

Case 1a $=$ Case 1 with $m_3 = 0$;

Case 1b $=$ Case 1 with one more mass (not $m_3$) equal to zero.

The cases 1,2,3,4 are independent, and we shall discuss them separately.

As in the paper \cite{DST}, we shall use general results on
asymptotic expansions of Feynman diagrams.
This subject was developed in several papers \cite{as-ex} (see also
\cite{Smirnov-book} for a review).

Before presenting these results in our case,
we would like to discuss some issues connected with this approach.
Let us first introduce the Taylor expansion operator
${\cal T}_k$. Its action on the denominator can be represented as
\be
\label{Taylor}
{\cal T}_{k} \; \frac{1}{{[(k-p)^2 - m^2]}^{\nu}}
= \sum_{j=0}^{\infty} \frac{(\nu)_j}{j!} \;
    \frac{(2 (kp) - k^2)^j}{[p^2 - m^2]^{\nu+j}}  \; ,
\ee
where
\be
\label{poch}
(\nu)_j \equiv \frac{\Gamma(\nu+j)}{\Gamma(\nu)}
\ee
is the Pochhammer symbol. It is understood that this operator
${\cal T}$ acts on the integrands before the loop integrations
are performed, and also that all the terms
with equal powers of $k$ on the r.h.s. should be collected together
(e.g. the terms with $k^2$ and $(kp)^2$ should be considered together).
Below we shall also need a generalization of formula (\ref{Taylor})
to the expansion in two or more momenta ($k$ and one or two
loop momenta). This generalization is straightforward: we can
expand in all these momenta at the same time.
We only have to remember that in this case the expansion
is in the total power of these momenta, because inserting
loop momenta in the numerator of the resulting massless
integral effectively produces higher powers of $k$.

Since the limit $k^2 \to 0$ is not regular for the cases
being considered, the true expansion cannot be obtained simply
by applying eq.~(\ref{Taylor}) to all denominators where the external
momentum $k$ occurs. If we did this, we would obtain
infrared singularities in the coefficients of the expansion,
even if the initial diagram were finite.
This is due to the fact that in all zero-threshold  cases
we always need to let the momentum $k$ go through at least
one of the massless lines. Nevertheless, it is possible
to correct this procedure by adding some extra terms.
The general theorem on asymptotic expansions tells us
how to construct this true expansion. Namely, the result
is given by the na\"{\i}ve Taylor expansion in small parameters
(in our case, ${\cal T}_k$) plus terms given by Taylor expansions
in a specific class of subgraphs (see below) which may also
contain expansions in some of the loop momenta.
So, this na\"{\i}ve Taylor series produces infrared singularities
while the extra series also involve ultraviolet ones,
due to the appearance of loop momenta in the numerator.

If the original diagram is finite, all the singularities should
cancel\footnote{If it contains singularities of a specific type,
those will survive.}.
Moreover, infrared and ultraviolet singularities should
cancel independently.
It is possible to write general formulae of asymptotic expansions
in such an explicitly finite form, but
such a procedure is more complicated.
However, if we use
dimensional regularization for both types of singularities,
some contributions correspond to massless tadpoles and vanish.
In this case, the formulae are much simpler, and we are going
to use this approach below. It might seem in the resulting
formulae that now infrared poles
in $\ep$ are cancelled by ultraviolet ones. The explanation is
that omitting massless tadpoles effectively leads to
a ``re-distribution'' of both types of singularities.
So, from a practical point of view it is more convenient to apply the
formulae where, formally, poles in $\ep$ corresponding
to infrared and ultraviolet singularities of the integrand are
mutually cancelled. Moreover, cancellation
of these poles for convergent diagrams will be a good check of our
calculational procedure.

Let us introduce the following notation: $\Gamma$ is the original
graph corresponding in our case to Fig.~1a, subgraphs of $\Gamma$
are denoted as $\gamma$, and the corresponding ``reduced graph''
$\Gamma/\gamma$ is obtained from $\Gamma$ by shrinking the subgraph
$\gamma$ to a point. Furthermore, $J_{\gamma}$ is the
dimensionally-regularized Feynman
integral with the denominators corresponding to a graph $\gamma$.
For example, $J_{\Gamma}$ corresponds to the integral (\ref{defJ}).
For our case,
the general theorem yields
\be
\label{theorem}
J_{\Gamma}  \begin{array}{c} \frac{}{}  \\
                    {\mbox{\Huge$\sim$}} \\ {}_{k^2 \to 0}
            \end{array}
\sum_{\gamma} J_{\Gamma / \gamma}
\circ {\cal T}_{k, \; q_{i}} J_{\gamma} ,
\ee
where the sum goes over all the subgraphs $\gamma$ which
(i) contain all the lines with large masses,
and (ii) are one-particle irreducible with respect to the
light (in our case, massless) lines.
${\cal T}_{k, \; q_{i}}$ is an operator (\ref{Taylor})
expanding the integrand in $k$ and the loop momenta $q_i$ which are
``external'' for a given $\gamma$.
The symbol ``$\circ$'' means that the polynomial in $q_i$,
which appears as a result of applying ${\cal T}$ to $J_{\gamma}$,
should be inserted in the numerator of the integrand
of $J_{\Gamma / \gamma}$. Note that although eq.~(\ref{theorem})
looks similar to the corresponding result for the large
momentum expansion, eq.~(2) of \cite{DST}, there is an essential
difference in the choice of subgraphs $\gamma$ contributing
to the expansion.

\begin{figure}[htb]
 \[
 \begin{array}{ll}
 \raisebox{10pt}{Case 1 } & \epsffile{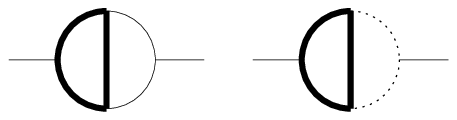}  \\
 \raisebox{10pt}{Case 1a} & \epsffile{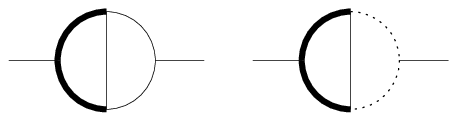} \\
 \raisebox{10pt}{Case 1b} & \epsffile{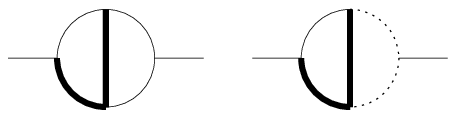} \\
 \raisebox{10pt}{Case 2 } & \epsffile{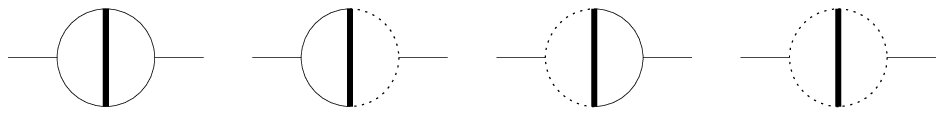}  \\
 \raisebox{10pt}{Case 3 } & \epsffile{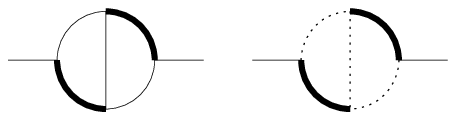}  \\
 \raisebox{10pt}{Case 4 } & \epsffile{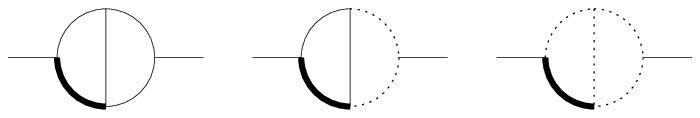}  \\
  \end{array}
  \]
      \caption[]{ The subgraphs $\gamma$ to be included in the
            sum~(\ref{theorem}).}
\label{fig:subgraphs}
\end{figure}

Now, let us consider the contributions to the sum (\ref{theorem})
for the different threshold configurations.
We shall use the numbering given in Fig.~1a to
indicate which lines are included in the subgraph:
for example, $\{134\}$ denotes the subgraph containing only the
``left'' triangle (lines 1,3,4) of Fig.~1a, $\{12345\} \equiv \Gamma$,
etc. These subgraphs are illustrated by Fig.~\ref{fig:subgraphs},
where bold lines correspond to massive propagators,
narrow lines to massless ones, and dotted lines
to the lines omitted in the subgraph (as compared with
the ``full'' graph $\Gamma$). These dotted lines correspond to
the reduced graphs $\Gamma/\gamma$. We do not list the vanishing
contributions containing massless tadpoles.

{\em Case 1}. One 2PT is zero:  $m_2=m_5=0$.
Two subgraphs contribute to the asymptotic expansion: \\
(i) $\gamma=\Gamma \; \;  \Rightarrow$
\be
\label{c1-1}
\int\int \mbox{d}^n p \; \mbox{d}^n q
\frac{1}{[(p-q)^2-m_3^2]^{\nu_3} [p^2-m_4^2]^{\nu_4} [q^2]^{\nu_5}}
{\cal T}_{k}
\frac{1}{[(k-p)^2-m_1^2]^{\nu_1} [(k-q)^2]^{\nu_2}} ;
\ee
(ii) $\gamma = \{ 134 \} \; \; \Rightarrow$
\be
\label{c1-2}
\int\int \mbox{d}^n p \; \mbox{d}^n q
\frac{1}{ [(k-q)^2]^{\nu_2}[q^2]^{\nu_5}}
{\cal T}_{k,q}
\frac{1}{[(k-p)^2-m_1^2]^{\nu_1}[(p-q)^2-m_3^2]^{\nu_3}
[p^2-m_4^2]^{\nu_4}} .
\ee
In cases 1a and 1b we have the same
contributing subgraphs (since the threshold configuration is
the same), and the only thing to change is to set $m_3=0$ (case 1a)
or $m_4=0$ (case 1b) in the formulae (\ref{c1-1})--(\ref{c1-2}).

{\em Case 2}. Two 2PT's are zero: $m_1 = m_2 = m_4 = m_5 = 0,
\; m_3 \equiv m $.
In this case, four subgraphs contribute: \\
(i) $\gamma=\Gamma \; \; \Rightarrow$
\be
\label{c2-1}
\int\int \mbox{d}^n p \; \mbox{d}^n q
\frac{1}{[(p-q)^2-m^2]^{\nu_3} [p^2]^{\nu_4} [q^2]^{\nu_5}}
{\cal T}_{k}
\frac{1}{[(k-p)^2]^{\nu_1} [(k-q)^2]^{\nu_2}} ;
\ee
(ii) $\gamma = \{ 134 \} \; \; \Rightarrow$
\be
\label{c2-2}
\int\int \mbox{d}^n p \; \mbox{d}^n q
\frac{1}{ [(k-q)^2]^{\nu_2}[q^2]^{\nu_5}}
{\cal T}_{k,q}
\frac{1}{[(k-p)^2]^{\nu_1}[(p-q)^2-m^2]^{\nu_3}
[p^2]^{\nu_4}} ;
\ee
(iii) $\gamma = \{ 235 \}$ contribution can be obtained from the previous
one by the permutation $1 \leftrightarrow 2, 4 \leftrightarrow 5$; \\
(iv) $\gamma = \{ 3 \} \; \; \Rightarrow$
\be
\label{c2-4}
\int\int \mbox{d}^n p \; \mbox{d}^n q
\frac{1}{[(k-p)^2]^{\nu_1}[(k-q)^2]^{\nu_2}[p^2]^{\nu_4}[q^2]^{\nu_5}}
{\cal T}_{p,q}
\frac{1}{[(p-q)^2-m^2]^{\nu_3}} .
\ee

{\em Case 3}. One 3PT is zero: $m_2=m_3=m_4=0$.
Two subgraphs contribute: \\
(i) $\gamma=\Gamma \; \; \Rightarrow$
\be
\label{c3-1}
\int\int \mbox{d}^n p \; \mbox{d}^n q
\frac{1}{[(p-q)^2]^{\nu_3} [p^2]^{\nu_4} [q^2-m_5^2]^{\nu_5}}
{\cal T}_{k}
\frac{1}{[(k-p)^2-m_1^2]^{\nu_1} [(k-q)^2]^{\nu_2}} ;
\ee
(ii) $\gamma = \{ 15 \} \; \; \Rightarrow$
\be
\label{c3-2}
\int\int \mbox{d}^n p \; \mbox{d}^n q
\frac{1}{[(k-q)^2]^{\nu_2} [(p-q)^2]^{\nu_3} [p^2]^{\nu_4}}
{\cal T}_{k,p,q}
\frac{1}{[(k-p)^2-m_1^2]^{\nu_1} [q^2-m_5^2]^{\nu_5}} .
\ee
%NB: probably, it is better to choose another set of loop momenta, for
%this contribution (e.g. $q \to k-q, \; p \to k-p$).

{\em Case 4}. One 2PT and one 3PT are zero: $m_2=m_3=m_4=m_5=0, \;
m_1 \equiv m$.
For this case, three subgraphs contribute: \\
(i) $\gamma=\Gamma \; \; \Rightarrow$
\be
\label{c4-1}
\int\int \mbox{d}^n p \; \mbox{d}^n q
\frac{1}{[(p-q)^2]^{\nu_3} [p^2]^{\nu_4} [q^2]^{\nu_5}}
{\cal T}_{k}
\frac{1}{[(k-p)^2-m^2]^{\nu_1} [(k-q)^2]^{\nu_2}} ;
\ee
(ii) $\gamma = \{ 134 \} \; \; \Rightarrow$
\be
\label{c4-2}
\int\int \mbox{d}^n p \; \mbox{d}^n q
\frac{1}{ [(k-q)^2]^{\nu_2}[q^2]^{\nu_5}}
{\cal T}_{k,q}
\frac{1}{[(k-p)^2-m^2]^{\nu_1}[(p-q)^2]^{\nu_3}
[p^2]^{\nu_4}} ;
\ee
(iii) $\gamma = \{ 1 \} \; \; \Rightarrow$
\be
\label{c4-3}
\int\int \mbox{d}^n p \; \mbox{d}^n q
\frac{1}{[(k-q)^2]^{\nu_2} [(p-q)^2]^{\nu_3} [p^2]^{\nu_4}
[q^2]^{\nu_5}}
{\cal T}_{k,p}
\frac{1}{[(k-p)^2-m^2]^{\nu_1}} .
\ee

Examination of eqs.~(\ref{c1-1})--(\ref{c4-3}) shows that,
after partial fractioning is performed wherever necessary,
the following types of contributions can occur in the expressions
for the coefficients of the zero-threshold expansion:\\
(a) two-loop vacuum diagrams with two (or one) massive and
    one (or two) massless lines;\\
(b) products of a one-loop massless diagram with external
    momentum $k$ and a one-loop massive tadpole;\\
(c) two-loop massless diagrams with one (or two) powers of
    the denominators being non-positive.\\
The algebra of contributions of type (a) is discussed
in Appendix A, which is based on a slightly modified algorithm of \cite{DT}.
The problem of numerators of massive integrals
occurring in types (a) and (b) has been studied in \cite{DST}
(see Appendix B of that paper), and we use the same formulae here.
For the numerators occurring in the integrals of
type (c), we have used a procedure based on a formula
in \cite{Davyd'91}, and we explain the details in Appendix B.

All the formulae~(\ref{c1-1})--(\ref{c4-3}), along with
eq.~(\ref{theorem}), can be used for any values of the
space-time dimension $n$ and the powers of
propagators $\nu_i$. In the next section, we shall consider
the results produced by the expansion (\ref{theorem}) for
some special cases.

\vspace{3mm}

% ---- \input{sec3}
\begin{center}
{\bf 3. Analytic results}
\end{center}

To get analytic results for the coefficients of the zero-threshold
expansion, we have written a computer program using
the {\sf REDUCE} system for analytical calculations \cite{Reduce}.
Like the algorithm itself, it works for any (integer) powers
of the propagators, and can be applied for both convergent
and divergent diagrams.

Many interesting physical applications require the calculation of
the diagram in Fig.~1a with unit powers of propagators,
the so-called ``master'' diagram. If all five $\nu_i$ are equal
to one (and $k^2 \neq 0$) the corresponding diagram is finite
as $n \to 4$, and we shall calculate the corresponding results
at $n=4$. The algorithm also makes it possible to consider
higher terms of the expansion in $\ep = (4-n)/2$, or even
results for arbitrary $n$.
The individual contributions to the expansion (\ref{c1-1})--(\ref{c4-3})
contain poles in $\ep$, which, for finite diagrams such as
this ``master'' diagram, cancel in the sum (\ref{theorem}).
This is one of the non-trivial checks
on the algorithm.

Let us write the corresponding
``master'' integral as
\be
\label{J-exp}
J(m_1,m_2,m_3,m_4,m_5; k)
\equiv J(1,1,1,1,1;m_1,m_2,m_3,m_4,m_5; k) =
% \equiv J(1,1,1,1,1;\{m_i\}; k) =
% J(m_1,m_2,m_3,m_4,m_5; k) =
- \pi^4 \sum_{j=0}^{\infty} C_j \; (k^2)^j ,
\ee
where, for zero-threshold configurations, the coefficients $C_j$ can
depend on logarithms of $k^2$,
\be
\label{C_j}
C_j = C_j^{(2)} \ln^2(-k^2) + C_j^{(1)} \ln(-k^2) + C_j^{(0)}.
\ee
Powers of $\ln(-k^2)$ higher than two cannot occur
in the cases considered, because the separate terms contributing
to the expansion can have at most double poles in $\ep$.

%======================== TABLE 1 =========================================
\begin{table}[t]
\begin{tabular}{|l|l|l|l|l|l|l|}
\hline
$j$& ${m0mm0}$ & ${m00m0}$ & ${m0m00}$ & ${00m00}$ & ${m000m}$ & ${m0000}$ \\
{} & (case 1)  & (case 1a) & (case 1b) & (case 2)  & (case 3)  & (case 4)  \\
\hline
\hline
{0} &
${\sfrac{1}{2} L}$ &
${L}$&
${L}$&
${-L^2 + 2 L}$&
${ }$&
${-\sfrac{1}{2} L^2 + 2 L}$\\
{}&
${-\sfrac{3}{2}}$ &
${-2}$&
${+\zet{2} - 4}$&
${-2\zet{2}}$&
${-2\zet{2}}$&
${-\zet{2} - 3}$\\
\hline
{1} &
${\sfrac{1}{24} L}$ &
${\sfrac{1}{12} L}$&
${\sfrac{1}{4} L}$&
${\sfrac{1}{2} L^2 - \sfrac{1}{2} L}$&
${\sfrac{1}{2} L }$&
${-\sfrac{1}{4} L^2 + \sfrac{1}{2} L}$\\
{}&
${-\sfrac{1}{16}}$ &
${+\sfrac{1}{12}}$&
${+\sfrac{1}{2} \zet{2} - \sfrac{13}{12}}$&
${+\zet{2} - \sfrac{3}{2}}$&
${-\zet{2} - \sfrac{1}{4}}$&
${-\sfrac{1}{2} \zet{2} - \sfrac{7}{8}}$\\
\hline
{2} &
${\sfrac{1}{180} L}$ &
${\sfrac{1}{90} L}$&
${\sfrac{1}{9} L}$&
${-\sfrac{1}{3} L^2 + \sfrac{2}{9} L}$&
${\sfrac{1}{3} L}$&
${-\sfrac{1}{6} L^2 + \sfrac{2}{9} L}$\\
{}&
${-\sfrac{1}{180}}$ &
${+\sfrac{23}{540}}$&
${+\sfrac{1}{3} \zet{2} - \sfrac{673}{1080}}$&
${-\sfrac{2}{3} \zet{2} + 1}$&
${-\sfrac{2}{3} \zet{2} + \sfrac{13}{36}}$&
${-\sfrac{1}{3} \zet{2} - \sfrac{11}{18}}$\\
\hline
{3} &
${\sfrac{1}{1120} L}$ &
${\sfrac{1}{560} L}$&
${\sfrac{1}{16} L}$&
${\sfrac{1}{4} L^2 - \sfrac{1}{8} L}$&
${\sfrac{5}{24} L}$&
${-\sfrac{1}{8} L^2 + \sfrac{1}{8} L}$\\
{}&
${-\sfrac{3}{4480}}$ &
${+\sfrac{17}{1120}}$&
${+\sfrac{1}{4} \zet{2} - \sfrac{1487}{3360}}$&
${+\sfrac{1}{2} \zet{2}  - \sfrac{115}{144}}$&
${-\sfrac{1}{2} \zet{2} + \sfrac{763}{1440}}$&
${-\sfrac{1}{4} \zet{2} - \sfrac{287}{576}}$\\
\hline
{4} &
${\sfrac{1}{6300} L}$ &
${\sfrac{1}{3150} L}$&
${\sfrac{1}{25} L}$&
${-\sfrac{1}{5} L^2 + \sfrac{2}{25} L}$&
${\sfrac{2}{15} L}$&
${-\sfrac{1}{10} L^2 + \sfrac{2}{25} L}$\\
{}&
${-\sfrac{1}{10500}}$ &
${+\sfrac{367}{63000}}$&
${+\sfrac{1}{5} \zet{2}- \sfrac{86939}{252000}}$&
${-\sfrac{2}{5} \zet{2}  + \sfrac{23}{36}}$&
${-\sfrac{2}{5} \zet{2} + \sfrac{13591}{25200}}$&
${-\sfrac{1}{5} \zet{2} - \sfrac{3841}{9000}}$\\
\hline
{5} &
${\sfrac{1}{33264} L}$ &
${\sfrac{1}{16632} L}$&
${\sfrac{1}{36} L}$&
${\sfrac{1}{6} L^2 - \sfrac{1}{18} L}$&
${\sfrac{4}{45} L}$&
${-\sfrac{1}{12} L^2 + \sfrac{1}{18} L}$\\
{}&
${-\sfrac{1}{66528}}$ &
${+\sfrac{2531}{997920}}$&
${+\sfrac{1}{6} \zet{2} - \sfrac{2828267}{9979200}}$&
${+\sfrac{1}{3} \zet{2} - \sfrac{973}{1800}}$&
${-\sfrac{1}{3} \zet{2} + \sfrac{75683}{151200}}$&
${-\sfrac{1}{6} \zet{2} - \sfrac{4049}{10800}}$\\
\hline
{6} &
${\sfrac{1}{168168} L}$ &
${\sfrac{1}{84084} L}$&
${\sfrac{1}{49} L}$&
${-\sfrac{1}{7} L^2 + \sfrac{2}{49} L}$&
${\sfrac{13}{210} L}$&
${-\sfrac{1}{14} L^2 + \sfrac{2}{49} L}$\\
{}&
${-\sfrac{1}{392392}}$ &
${+\sfrac{4903}{3923920}}$&
${+\sfrac{1}{7} \zet{2} \! - \! \sfrac{85046849}{353152800}}$&
${-\sfrac{2}{7} \zet{2} \! + \! \sfrac{139}{300}}$&
${-\sfrac{2}{7} \zet{2} \! + \! \sfrac{2629903}{5821200}}$&
${-\sfrac{1}{7} \zet{2} \! - \! \sfrac{206741}{617400}}$\\
%\hline
%{7} &
%${\sfrac{1}{823680} L}$ &
%${\sfrac{1}{411840} L}$&
%${}$&
%${\sfrac{1}{8} L^2 - \sfrac{1}{32} L}$&
%${}$&
%${}$\\
%{}&
%${-\sfrac{1}{2196480}}$ &
%${+\sfrac{39379}{57657600}}$&
%${}$&
%${+\sfrac{1}{4} \zet{2} \! - \! \sfrac{191833}{470400}}$&
%${}$&
%${}$\\
\hline \hline
{} &
compared &
compared &
$L$-terms &
compared &
$L$-terms &
compared \\
{} &
with \cite{BFT}, &
with \cite{ST}, &
checked &
with \cite{ST}, &
checked &
with \cite{BFT}, \\
{} &
eq.~(23) &
eq.~(105) &
{} &
eq.~(104) &
{} &
eqs.~(13)-(14) \\
\hline
\end{tabular}
\caption{Coefficients of zero-threshold expansion for the cases
with only one mass.
%( $L \eqiuv \ln(-k^2/m^2)$ )
}
\end{table}
%================================================================

For cases when we have only one massive parameter
$m$ (all non-zero masses are equal), a number of exact
results are known \cite{Broadh,BFT,ST}; for some of them, closed
expressions for the coefficients of the expansion are also
available. They can be represented
in terms of polylogarithms $\mbox{Li}_N$ with $N \leq 3$.
While applying the algorithm described in Section~2
we have considered all possibilities to have only one
mass parameter in zero-threshold configurations.
They correspond to the diagrams in the first column of
Fig.~3 with all non-zero masses equal.
The results for the first few coefficients of the expansion
are presented in Table~1, where the first line gives
the mass arguments of $J$, and the dimensionless coefficients
(and other notations) are defined by
\be
\label{one-mass}
J =
- \frac{\pi^4}{m^2}
\sum_{j=0}^{\infty} c_j \; \left( \frac{k^2}{m^2} \right)^j,\hspace{1cm}
%z \equiv \frac{k^2}{m^2}, \hspace{1cm}
L \equiv \ln\left(- \frac{k^2}{m^2} \right) , \hspace{1cm}
\zeta_2 \equiv \zeta(2) = \frac{\pi^2}{6} .
\ee
So, in Table~1 the following coefficients $c_j \; (j=0,\ldots,6)$ are
presented,
\be
c_j \equiv (m^2)^{j+1} C_j
= c_j^{(2)} L^2 + c_j^{(1)} L + c_j^{(0)} ,
\ee
and they agree with exact results wherever
available (see the last line of the table). For
cases 1b and 3 exact results were not available, but we
checked the logarithmic terms by use of the dispersion
relations technique.

In cases 2 and 4, we have only one massive line (one mass
parameter), and the expansion (\ref{one-mass}) is sufficient.
However, for other zero-threshold configurations we can have
two (cases 1a, 1b, 3) or even three (general case 1) different masses.

Let us start by considering case~1 with three different masses.
We have obtained the following results for the lowest
coefficients of the expansion (\ref{J-exp}):
%The coefficient at $(k^2)^0$ is
\bea
\label{C0-case1-3m}
C_0 =
\frac{1}{4 (m_1^2 - m_3^2)(m_4^2 - m_3^2) (m_1^2 - m_4^2)} \;
\hspace{80mm}
\nonumber \\[2mm]
\times
\left\{ 2 \left(\ln\left(-\frac{k^2}{m_1^2}\right)
              + \ln\left(-\frac{k^2}{m_4^2}\right) - 4 \right) \;
        \left( m_1^2 (m_4^2 - m_3^2) \ln\frac{m_1^2}{m_3^2}
             - m_4^2 (m_1^2 - m_3^2) \ln\frac{m_4^2}{m_3^2} \right)
\right.
\nonumber \\[2mm]
       + (m_1^2 - m_3^2) (m_4^2 - m_3^2)
          \ln\frac{m_1^2}{m_4^2}
          \left( \ln\frac{m_1^2}{m_3^2} + \ln\frac{m_4^2}{m_3^2} \right)
       - 2 m_3^2 (m_1^2 - m_4^2)
          \ln\frac{m_1^2}{m_3^2} \; \ln\frac{m_4^2}{m_3^2}
\nonumber \\[2mm]
\left.
       + 2 (m_1^2 + m_3^2) (m_4^2 - m_3^2) {\cal{H}}(m_1^2, m_3^2)
       - 2 (m_4^2 + m_3^2) (m_1^2 - m_3^2) {\cal{H}}(m_4^2, m_3^2)
\frac{}{} \right\} ,
\hspace{6mm}
\eea
\bea
\label{C1-case1-3m}
C_1 =
\frac{1}{8 (m_1^2 - m_3^2)^3 (m_4^2 - m_3^2)^3 (m_1^2 - m_4^2)^3} \;
\hspace{76mm}
\nonumber \\[2mm]
\times
\left\{ 2 (m_1^2 - m_3^2) (m_4^2 - m_3^2) \;
          \left(\ln\left(-\frac{k^2}{m_1^2}\right)
              + \ln\left(-\frac{k^2}{m_4^2}\right) - 4 \right) \;
\right.
\hspace{45mm}
\nonumber \\[2mm]
\times
\left( m_1^2 (m_4^2 - m_3^2)^2 (m_1^2 (m_1^2 + m_4^2) - 2 m_4^2 m_3^2)
           \ln\frac{m_1^2}{m_3^2}
\right.
\hspace{40mm}
\nonumber \\[2mm]
     - m_4^2 (m_1^2 - m_3^2)^2 (m_4^2 (m_1^2 + m_4^2) - 2 m_1^2 m_3^2)
           \ln\frac{m_4^2}{m_3^2}
\hspace{37mm}
\nonumber \\[2mm]
\left.
+  (m_1^2 - m_3^2) (m_4^2 - m_3^2) (m_1^2 - m_4^2)
      (m_3^2 (m_1^2 + m_4^2) - 2 m_1^2 m_4^2) \;
\frac{}{} \right)
\hspace{10mm}
\nonumber \\[2mm]
 + (m_1^2 - m_3^2)^2 (m_4^2 - m_3^2)^2 (m_1^2 + m_4^2) \;
   \left( \ln\frac{m_1^2}{m_3^2} + \ln\frac{m_4^2}{m_3^2} \right)
\hspace{50mm}
\nonumber \\[2mm]
\times
\left( (m_1^2 - m_3^2) (m_4^2 - m_3^2) \ln\frac{m_1^2}{m_4^2}
       + 2 m_3^2 (m_1^2 - m_4^2)
\right)
\hspace{37mm}
\nonumber \\[2mm]
 + 2 m_3^4 (m_1^2 - m_3^2) (m_4^2 - m_3^2) (m_1^2 - m_4^2)^3 \;
   \ln\frac{m_1^2}{m_3^2} \; \ln\frac{m_4^2}{m_3^2}
\hspace{56mm}
\nonumber \\[2mm]
 + 4 m_3^2 (m_1^2 - m_4^2)
\left( (m_1^2 - m_3^2) (m_4^2 - m_3^2) - (m_1^2 - m_4^2)^2 \right)
\hspace{50mm}
\nonumber \\[2mm]
\times \left( m_1^2 (m_4^2 - m_3^2)^2  \ln\frac{m_1^2}{m_3^2}
            + m_4^2 (m_1^2 - m_3^2)^2  \ln\frac{m_4^2}{m_3^2}
\right)
\hspace{30mm}
\nonumber \\[2mm]
 + 2 (m_1^2 - m_3^2) (m_4^2 - m_3^2) (m_1^2 - m_4^2)^3
     (m_1^2 m_3^2 + m_4^2 m_3^2 + m_1^2 m_4^2 - 3 m_3^4)
\hspace{28mm}
\nonumber \\[2mm]
+ 2 (m_1^2 - m_3^2) (m_4^2 - m_3^2)^3
  \left( (m_1^2 + m_4^2)(m_1^4 - m_3^4) + 2 m_1^2 m_3^2 (m_1^2 - m_4^2)
  \right) \; {\cal{H}}(m_1^2, m_3^2)
\hspace{5mm}
\nonumber \\[2mm]
\left.
- 2 (m_1^2 - m_3^2)^3 (m_4^2 - m_3^2)
  \left( (m_1^2 + m_4^2)(m_4^4 - m_3^4) - 2 m_4^2 m_3^2 (m_1^2 - m_4^2)
  \right) \; {\cal{H}}(m_4^2, m_3^2)
\frac{}{} \right\} .
\eea
In these formulae,
${\cal{H}}$ is a dimensionless function which is
defined as
\be
\label{defH}
{\cal{H}}(m_1^2, m_2^2) = 2\Li{2}{1- \frac{m_1^2}{m_2^2}}
               + \frac{1}{2} \ln^2\left(\frac{m_1^2}{m_2^2}\right).
\ee
It is easy to see that the function ${\cal{H}}$ is antisymmetric
in its arguments and therefore it vanishes when they are equal.
This function is connected with the
finite part of the two-loop vacuum integral with one massless and
two massive lines (for details of this definition, see Appendix A).
We have also obtained higher coefficients for this case (up to $C_3$),
but do not present them here because they are more cumbersome.

If two of the masses are equal, $m_1=m_4\equiv m$,
eqs.~(\ref{C0-case1-3m})--(\ref{C1-case1-3m}) can be simplified to
\bea
\label{C0-case1-2m}
C_0 =
\frac{1}{2 (m^2 - m_3^2)^2}
\left\{ 2\left( m^2 - m_3^2 - m_3^2 \ln\frac{m^2}{m_3^2} \right) \;
        \ln\left(-\frac{k^2}{m^2}\right)
\right.
\hspace{9mm}
\nonumber \\[2mm]
\left.
- m_3^2 \ln^2\frac{m^2}{m_3^2} - 4 (m^2 - m_3^2)
- 2 m_3^2 \; {\cal{H}}(m^2, m_3^2)
\right\} ,
\eea
\bea
\label{C1-case1-2m}
C_1 =
\frac{1}{12 m^2 (m^2 \! - \! m_3^2)^4}
\left\{
\left( 6 m^2 m_3^4 \ln\frac{m^2}{m_3^2}
       + (m^2 \! - \! m_3^2)(m^4 \! - \! 5 m^2 m_3^2 \!
                                           - \! 2 m_3^4) \right)
      \ln\left(-\frac{k^2}{m^2}\right)
\right.
\nonumber \\[2mm]
+ 3 m^2 m_3^4 \ln^2\frac{m^2}{m_3^2}
- 2 m_3^2 (5 m^4 - m_3^4) \ln\frac{m^2}{m_3^2}
\hspace{35mm}
\nonumber \\[2mm]
\left.
+ (m^2 - m_3^2) (m^4 + 17 m^2 m_3^2 + 2 m_3^4)
+ 6 m^2 m_3^4 \; {\cal{H}}(m^2, m_3^2)
\right\} .
\eea
In this case, we have got the coefficients up to $C_4$.

In another limit, $m_3 \to 0$ (this corresponds to the case~1a),
we get for the lowest coefficients
\be
\label{C0-case1a-2m}
C_0 =
\frac{1}{2 (m_1^2 - m_4^2)} \; \ln\frac{m_1^2}{m_4^2} \;
\left\{ \ln\left(-\frac{k^2}{m_1^2}\right)
      + \ln\left(-\frac{k^2}{m_4^2}\right) -4 \right\} ,
\ee
\bea
\label{C1-case1a-2m}
C_1 = \frac{1}{4 m_1^2 m_4^2 (m_1^2 - m_4^2)^3} \;
\left\{ m_1^2 m_4^2 (m_1^2 + m_4^2)
        \left(\ln\left(-\frac{k^2}{m_1^2}\right)
            + \ln\left(-\frac{k^2}{m_4^2}\right) - 4 \right) \;
        \ln\frac{m_1^2}{m_4^2}
\right. \hspace{2mm}
\nonumber \\[2mm]
\left.
       - 2 m_1^2 m_4^2 (m_1^2 - m_4^2)
        \left(\ln\left(-\frac{k^2}{m_1^2}\right)
            + \ln\left(-\frac{k^2}{m_4^2}\right) - 4 \right)
       + (m_1^2 - m_4^2)^3
\right\} ,
\eea
\bea
\label{C2-case1a-2m}
C_2 =
\frac{1}{36 m_1^4 m_4^4 (m_1^2 - m_4^2)^5} \;
\hspace{100mm}
\nonumber \\[2mm]
\times
\left\{ 6 m_1^4 m_4^4 (m_1^4 + 4 m_1^2 m_4^2 + m_4^4)
        \left(\ln\left(-\frac{k^2}{m_1^2}\right)
            + \ln\left(-\frac{k^2}{m_4^2}\right) -4 \right) \;
        \ln\frac{m_1^2}{m_4^2}
\right.
\hspace{20mm}
\nonumber \\[2mm]
       - 18 m_1^4 m_4^4 (m_1^4 - m_4^4)
        \left(\ln\left(-\frac{k^2}{m_1^2}\right)
            + \ln\left(-\frac{k^2}{m_4^2}\right) -4 \right) \;
\hspace{40mm}
\nonumber \\[2mm]
\left.
       - 2 m_1^4 m_4^4 (m_1^2 - m_4^2)^2 \;
           \ln\frac{m_1^2}{m_4^2}
       + (m_1^2 - m_4^2)^3 \; (m_1^2 + m_4^2) \;
          (m_1^4 - m_1^2 m_4^2 + m_4^4)
\right\} .
\eea
For this case (with two different masses), the exact result was also
available in \cite{ST}, and we have successfully compared our coefficients
with this exact expression.

Now, let us present the results for the coefficients (\ref{C_j})
for case 3
($m_2=m_3=m_4=0$), when $m_1$ and $m_5$ are different:
\be
\label{C0-case3-2m}
C_0 =
\frac{1}{4 m_1^2 m_5^2}
\left\{ - (m_1^2 + m_5^2)
       \left( \ln^2{\frac{m_1^2}{m_5^2}} + \frac{2 \pi^2}{3} \right)
        - 2 (m_1^2 - m_5^2)\; {\cal{H}}(m_1^2, m_5^2)
\right\} ,
\ee
\bea
\label{C1-case3-2m}
C_1 =
\frac{1}{8 m_1^4 m_5^4 (m_1^2 - m_5^2)}
\left\{ 2 m_1^2 m_5^2 (m_1^2 - m_5^2)
        \left( \ln\left(-\frac{k^2}{m_1^2}\right)
            + \ln\left(-\frac{k^2}{m_5^2}\right) - 3 \right)
\right.
\hspace{17mm}
\nonumber \\[2mm]
- (m_1^2 - m_5^2) (m_1^4 + m_5^4)
  \left( \ln^2{\frac{m_1^2}{m_5^2}} + \frac{2 \pi^2}{3} \right)
+ 2 m_1^2 m_5^2 (m_1^2 + m_5^2) \ln{\frac{m_1^2}{m_5^2}}
\nonumber \\[2mm]
\left. \frac{}{}
- 2 (m_1^2 - m_5^2)^2 \; (m_1^2 + m_5^2) \; {\cal{H}}(m_1^2, m_5^2)
\right\}  ,
\hspace{15mm}
\eea
\bea
\label{C2-case3-2m}
C_2 =
\frac{1}{36 m_1^6 m_5^6 (m_1^2 - m_5^2)^3}
\hspace{92mm}
\nonumber \\[2mm]
\times
\left\{ 3 m_1^2 m_5^2 (m_1^2 - m_5^2)^3 (m_1^2 + m_5^2)
        \left( \ln\left(-\frac{k^2}{m_1^2}\right)
            + \ln\left(-\frac{k^2}{m_5^2}\right) \right)
\right.
\hspace{10mm}
\nonumber \\[2mm]
- 3 (m_1^2 - m_5^2)^3 (m_1^2 + m_5^2) (m_1^4 - m_1^2 m_5^2 + m_5^4)
    \left( \ln^2{\frac{m_1^2}{m_5^2}} + \frac{2 \pi^2}{3} \right)
\nonumber \\[2mm]
+ 3 m_1^2 m_5^2 \left( 3 (m_1^2 - m_5^2)^4
                     + 4 m_1^2 m_5^2 (m_1^2 - m_5^2)^2
                          + 2 m_1^4 m_5^4 \right)
\ln{\frac{m_1^2}{m_5^2}}
\nonumber \\[2mm]
+ m_1^2 m_5^2 (m_1^2 - m_5^2) (m_1^2 + m_5^2)
 \left( (m_1^2 - m_5^2)^2 - 3 m_1^2 m_5^2 \right)
\nonumber \\[2mm]
\left.
- 6 (m_1^2 - m_5^2)^4 (m_1^4 + m_1^2 m_5^2 + m_5^4) \;
{\cal{H}}(m_1^2, m_5^2)
\frac{}{} \right\} .
\eea
Again, we do not present higher coefficients because they are more
cumbersome. For this case, we have calculated them up to $C_5$.

For all cases with different masses considered in this
section (except eqs.~(\ref{C0-case1a-2m})--(\ref{C2-case1a-2m})),
the exact results were not available. Nevertheless, we have
checked all the terms
containing $\ln(-k^2)$ by use of dispersion relations.
These logarithmic terms completely define
the imaginary part of the results, which is non-zero
for $k^2 > 0$.

To see that the algorithm also works correctly for cases
when the original integral is divergent, we considered an
example corresponding to the diagram in Fig.~1b, where the
first and the fourth lines were taken to be massive (with equal
masses) while the other three lines were massless. Effectively,
this example corresponds to case~3 (one zero-3PT) with
$\nu_1=2, \nu_2 = \nu_3 = \nu_4 = 1, \nu_5 = 0$.
The exact result for this case was presented in \cite{ST} (eq.~(102)),
and we have checked that our approach produces the correct coefficients
for both the divergent and finite parts (we have calculated terms
up to $C_6$).

Another possibility is to compare the obtained results with
programs based on numerical integration. This will also illustrate
the radius of convergence of the expansion. This comparison
is considered in the next section.

\vspace{3mm}

% ---- \input{sec4}
\begin{center}
{\bf 4. Numerical results}
\end{center}

In order to demonstrate the use of the zero-threshold expansion
as a way to obtain approximate numerical results for self-energy
diagrams, we shall discuss a number of mass configurations
corresponding to diagrams that occur in the Standard Model.
For this purpose, we take the masses of the $W$ and $Z$ bosons,
and of the top and bottom quarks to be:
\be
 M_W = \mbox{80 GeV}, \;\;
 M_Z = \mbox{91 GeV}, \;\;
 m_b = \mbox{5 GeV},  \;\;
 m_t = \mbox{174 GeV}.
\ee

\begin{figure}[htbp]
 \centerline{\epsffile{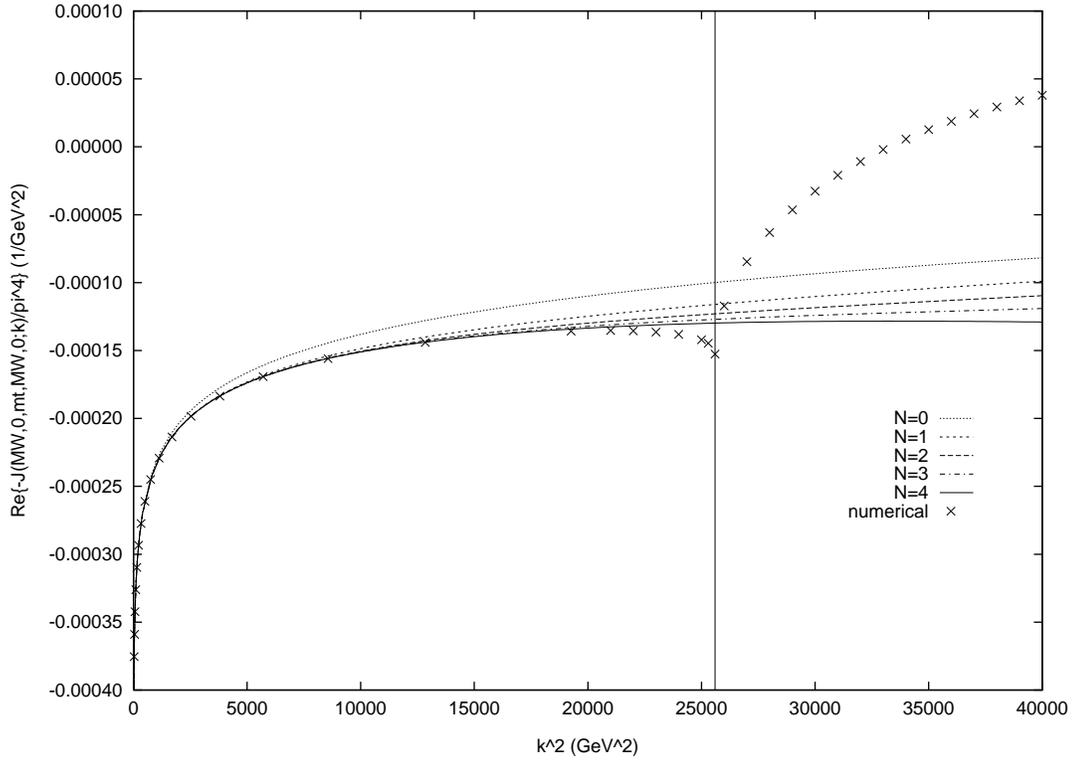}}
 \vspace{8mm}
 \centerline{\epsffile{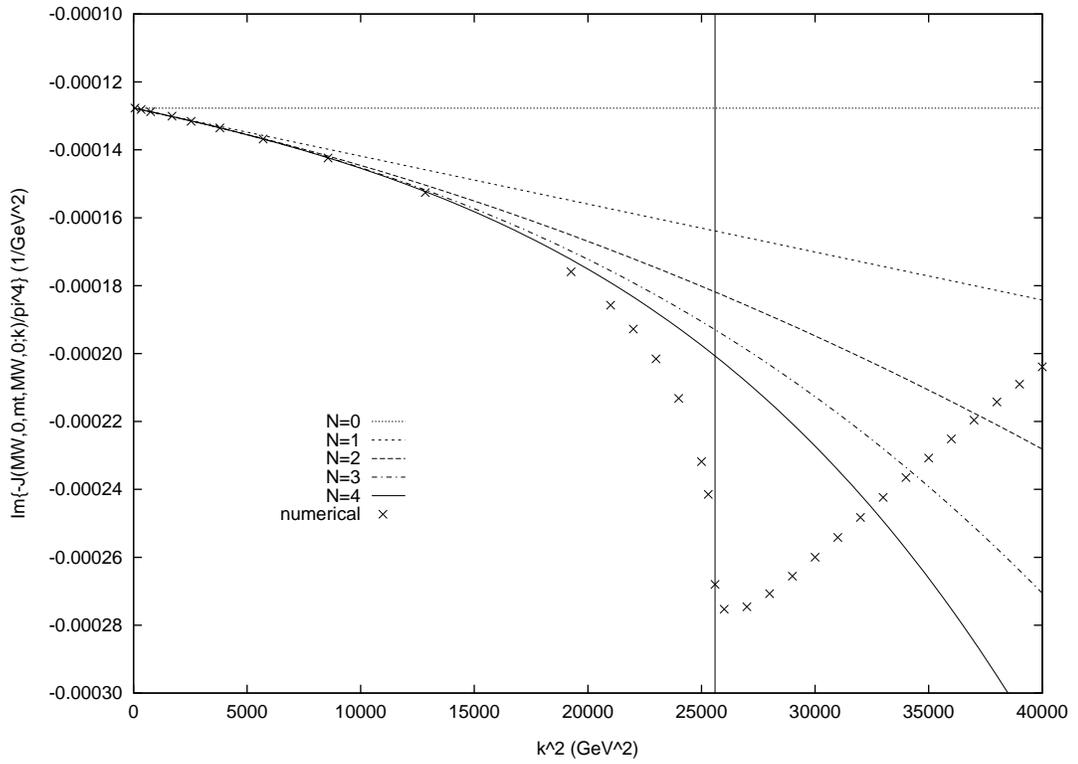}}
 \vspace{3mm}
 \caption[]{ The real and imaginary parts of $J(M_W,0,m_t,M_W,0;k)$.}
 \label{fig:W0tW0k}
\end{figure}

The first example is a diagram containing a top-bottom loop to
which two $W$ bosons are attached, that contributes to the
self-energy of the photon and the $Z$ boson. The corresponding
scalar integral is
\be
\label{eq:WbtWbk}
  J(M_W,m_b,m_t,M_W,m_b;k) \, .
\ee
Let us now neglect $m_b$ (below, we shall see that this is
reasonable when $k^2 \gg m_b^2$), and consider instead
\be
\label{eq:W0tW0k}
  J(M_W,0,m_t,M_W,0;k) \, .
\ee
It has one zero two-particle threshold and therefore it
belongs to case 1. In Fig.~\ref{fig:W0tW0k}
the approximations defined by
\be
J^{(N)} = - \pi^4 \sum_{j=0}^{N} C_j \; (k^2)^j
\ee
are shown as curves, and, for comparison, values of $J$ obtained by
numerical integration \cite{Kreimer} are shown as crosses.
The position of the lowest non-zero threshold, at $k^2=4M_W^2$
in this example, is marked by a vertical line.
In the real part, we clearly see the logarithmic singularity
as $k^2 \to 0$. This is the region where, by construction,
the asymptotic expansion works extremely well. For instance,
at $k^2=5000\mbox{ GeV}^2$, the $N=2$ approximation is already just
as good as the numerical integration, which is accurate to
4 digits, for both the real and imaginary parts.
However, the expansion converges all the way up to the
first non-zero threshold at $k^2=25600\mbox{ GeV}^2$. At that
point, the real parts of $J^{(4)}$ and of $J$ differ by about
15\%, and the imaginary parts by 25\%.

We also examined the situation for $k^2<0$, where there are no
physical thresholds and the imaginary part
is zero. At small $k^2$, the behaviour of the real part for negative
$k^2$ is quite
similar to that for positive $k^2$, because it is defined by the
$\ln |k^2|$ term in both cases. Moreover, because the signs of the
terms alternate for $k^2<0$, the convergence of the asymptotic
expansion is even a little better than for $k^2>0$.

\begin{figure}[htbp]
 \centerline{\epsffile{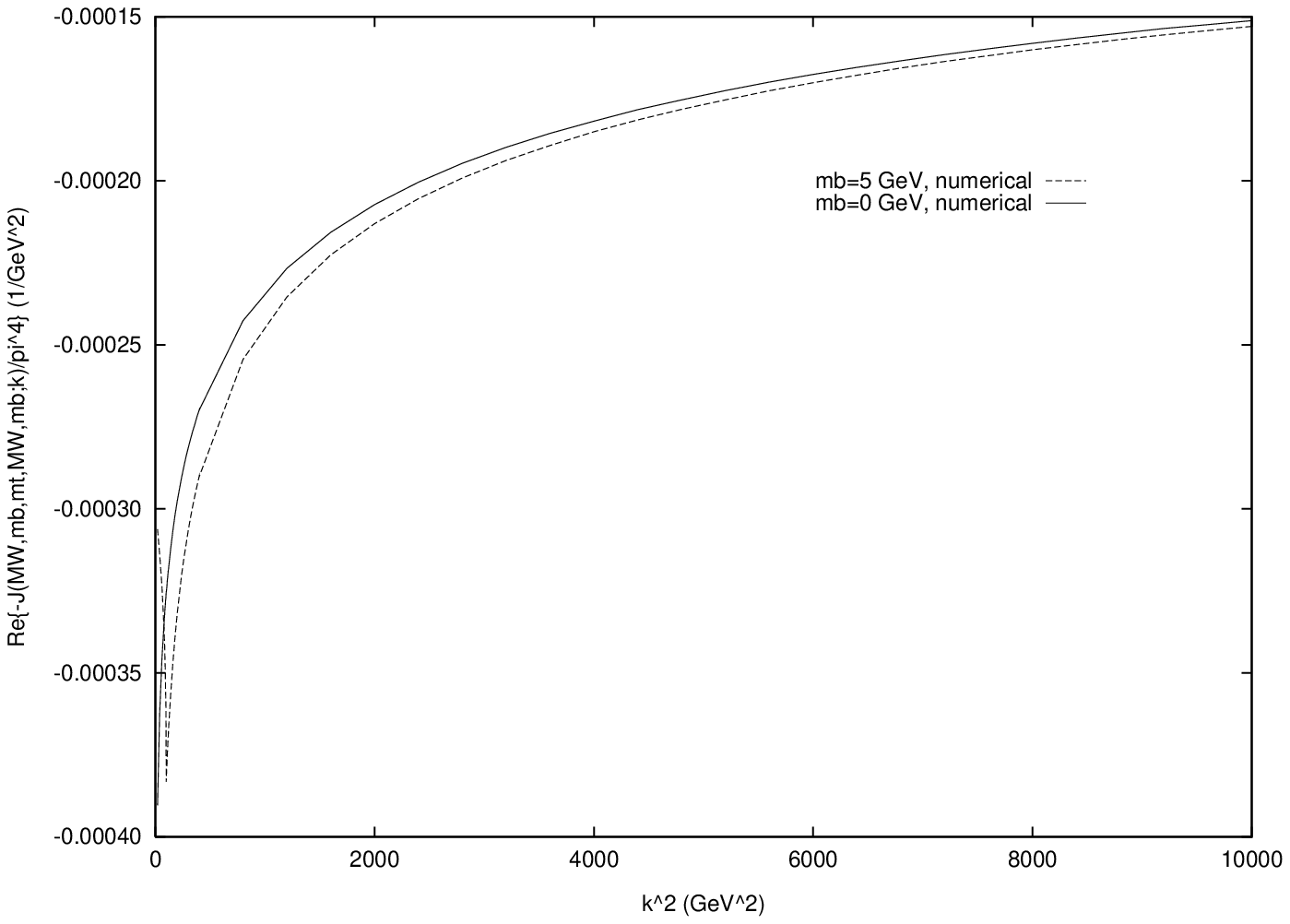}}
 \vspace{8mm}
 \centerline{\epsffile{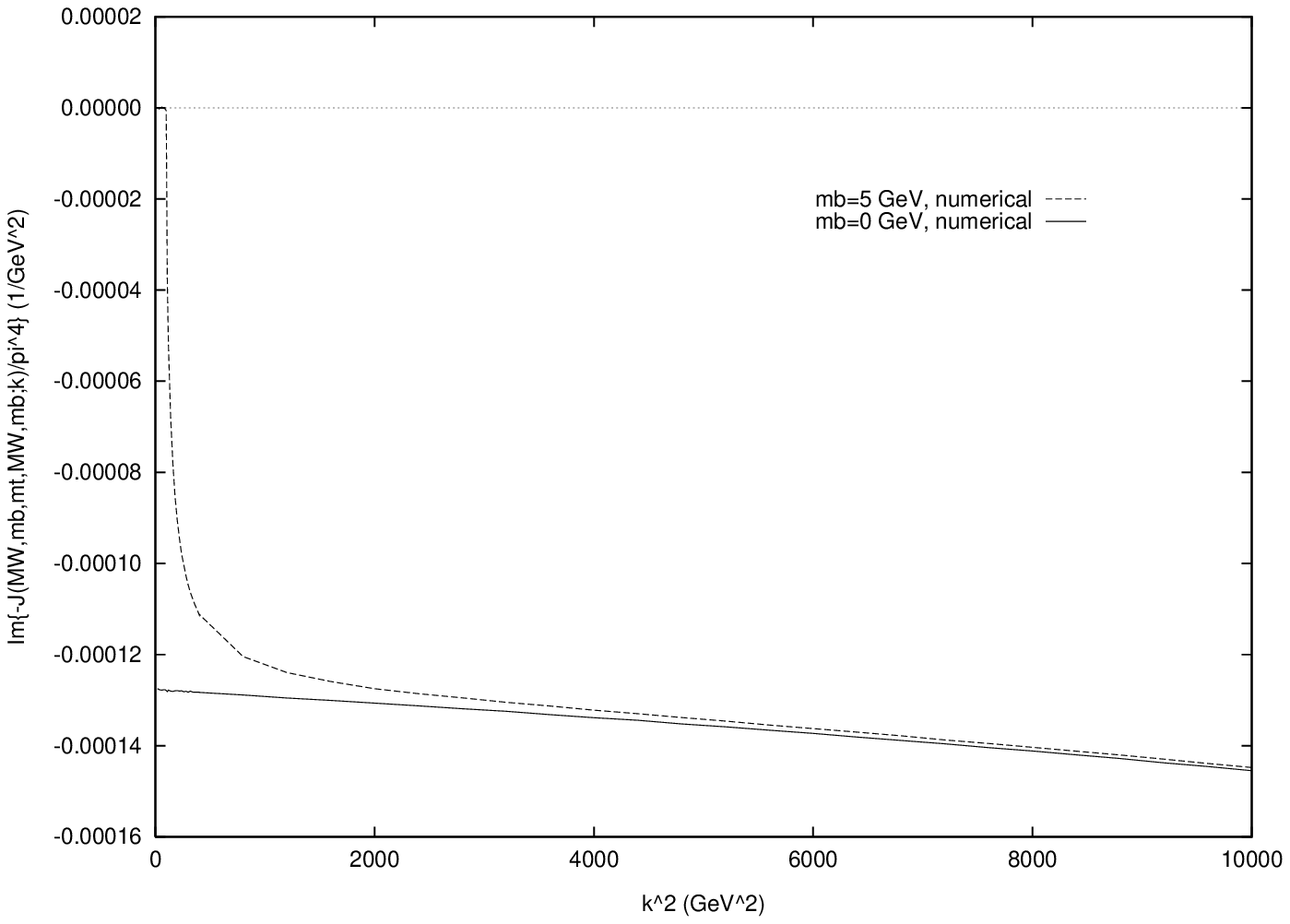}}
 \vspace{3mm}
 \caption[]{ The real and imaginary parts of
             $J(M_W,m_b,m_t,M_W,m_b;k)$
             and \\  $J(M_W,0,m_t,M_W,0;k)$.}
 \label{fig:mbeffect}
\end{figure}

In Fig.~\ref{fig:mbeffect},
values of the integrals (\ref{eq:WbtWbk}) (dashed lines)
and (\ref{eq:W0tW0k}) (solid lines) are plotted
to show the effect of neglecting $m_b$.
At low $k^2$, their behaviour is obviously very different,
the latter having a singularity at $k^2=0$, whereas the
former has one at $k^2=4m_b^2$, but as $k^2$ increases
above $4m_b^2$, they soon come close together.

For momenta close to the $Z$ mass shell,
$k^2\approx8300\mbox{ GeV}^2$, the error we make by
neglecting $m_b$ is of the order of 1\%.
The truncation error of the asymptotic expansion is
already less than that if we use $J^{(2)}$.

The second diagram we studied contains a top-bottom loop
with a $W$ boson going across the inside. Neglecting
$m_b$ as before, it corresponds to
\be
\label{eq:t0Wt0k}
  J(m_t,0,M_W,m_t,0;k)\, .
\ee
This mass configuration also belongs to case 1.
Below the first non-zero threshold, the behaviour of
(\ref{eq:t0Wt0k}) is
qualitatively very similar to the previous example,
but because this threshold is now a three-particle
threshold, it is different above the threshold.
In this case the truncation error of $J^{(4)}$ on the
threshold is only about 2\% in the real part and
0.4\% in the imaginary part.

\begin{figure}[htbp]
 \centerline{\epsffile{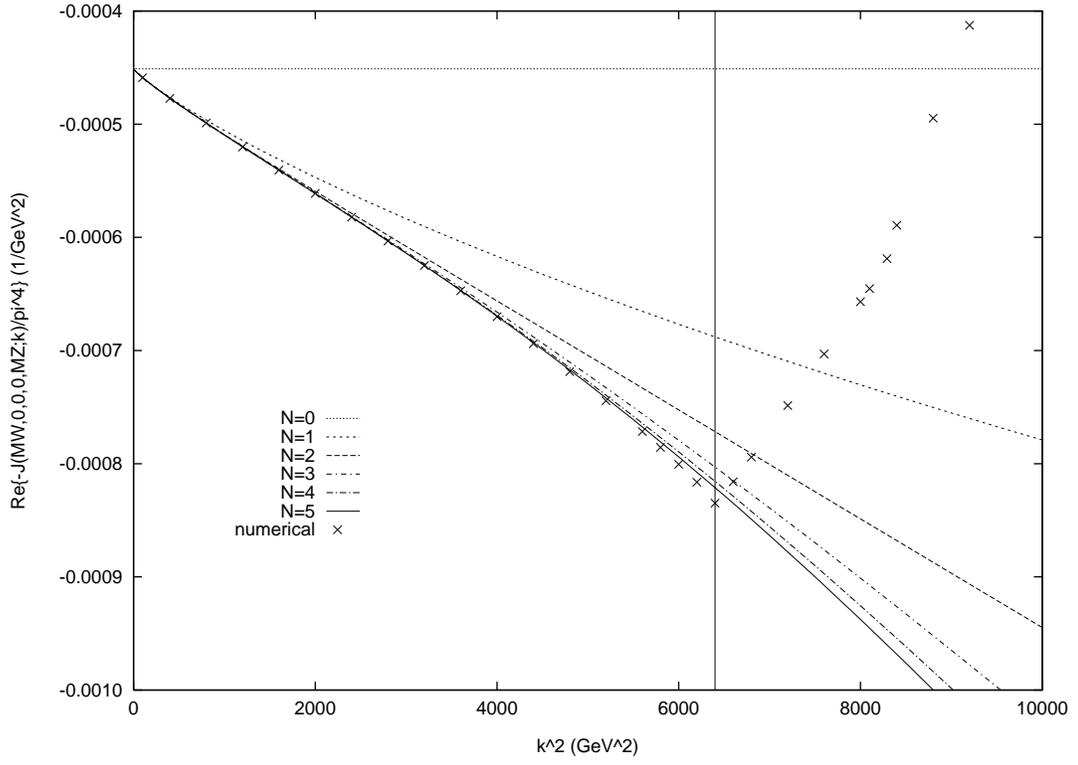}}
 \vspace{8mm}
 \centerline{\epsffile{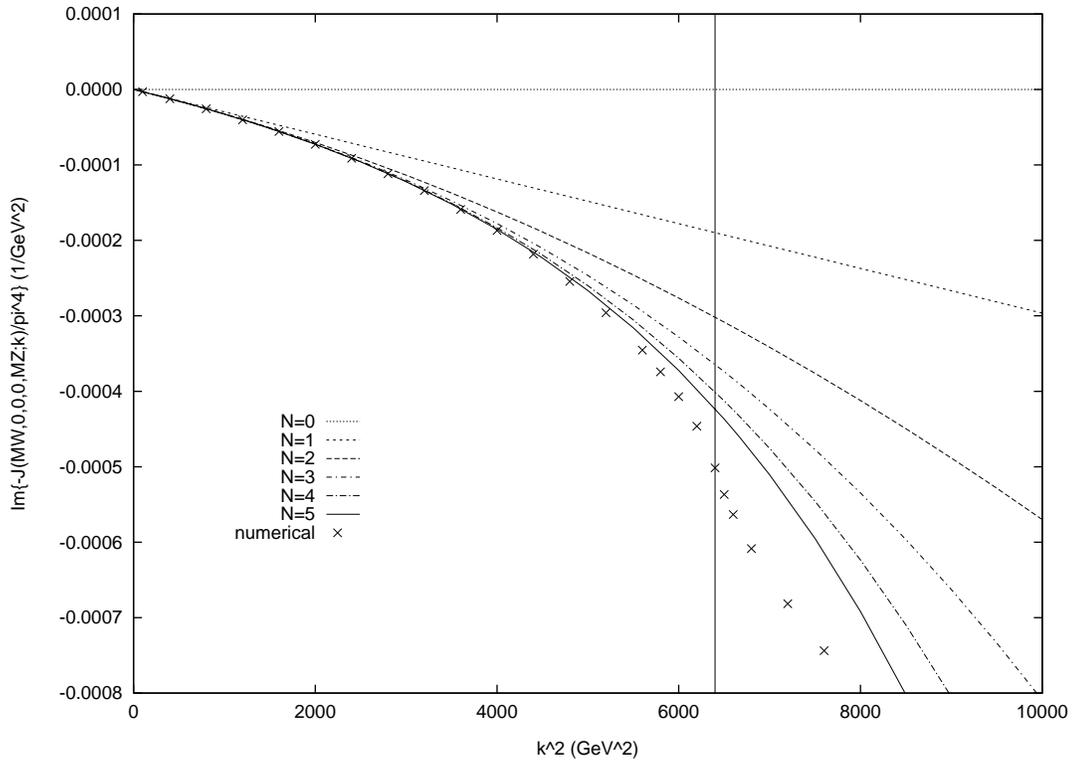}}
 \vspace{3mm}
 \caption[]{ The real and imaginary parts of $J(M_W,0,0,0,M_Z,k)$.}
 \label{fig:W000Zk}
\end{figure}

Finally, we study an example of case 3,
\be
\label{eq:W000Zk}
  J(M_W,0,0,0,M_Z;k) \, ,
\ee
which corresponds to diagrams that contribute to,
e.g., the electron or muon self-energy, if $m_e$
or $m_{\mu}$ is neglected with respect to
$k^2$. The results are shown in
Fig.~\ref{fig:W000Zk}.
The position of the first non-zero threshold, at
$k^2=M_W^2$, is again indicated by a vertical line.
For this case, we also checked that the numerical values
(shown as crosses) agree with the results of other
numerical programs \cite{Kreimer3,Japan}.

Contrary to case 1, the first term of the expansion,
$C_0$, given by eq.~(\ref{C0-case3-2m}), does not depend
on $\ln(-k^2)$, and therefore, the integral has a
finite limit as $k^2 \to 0$. At the threshold,
where the truncation errors are largest, the errors
in the real and imaginary parts of $J^{(5)}$ are approximately
2\% and 16\%, respectively.

In all the examples, the zero-threshold expansion provides approximations
that are at least as accurate as numerical integration in a large
part of the region of convergence, and can be evaluated much faster.

\vspace{3mm}

% ---- \input{conc.tex}
\begin{center}
{\bf 5. Conclusions}
\end{center}

In this paper we have studied all possibilities for a two-loop
self-energy diagram to have its lowest physical threshold at zero external
momentum squared. Due to specific infrared singularities at $k^2=0$,
constructing the small momentum expansion is more
complicated than when all the thresholds are non-vanishing. Namely,
instead of a regular Taylor expansion, we get a cut along the positive
$k^2$ axis, starting with a branchpoint at $k^2=0$. As a
consequence, our results in four dimensions contain $\ln(-k^2)$ terms, or
even $\ln^2(-k^2)$ terms if two physical thresholds vanish. So,
technically this zero-threshold expansion is closer to the large momentum
expansion \cite{DST} than to a regular small momentum expansion \cite{DT}.

For special cases, we have compared our results with known analytic ones.
For some more complicated cases corresponding to
diagrams with different masses occurring in the Standard Model, we made a
comparison with the results of a numerical integration program based on
the algorithm \cite{Kreimer} (for one case, we have also compared with
\cite{Japan}). It shows that the zero-threshold expansion converges up to
the first non-zero threshold, and that, unless $k^2$ is very close to
the threshold, only a few terms are needed to obtain accurate results.
This comparison can also be considered as a check of the
numerical programs.

The algorithm constructed in the present paper covers the
only ``hole'' remaining in the small momentum expansion of two-loop
self-energy diagrams. Together with \cite{DT}, it solves this
problem completely. The large momentum expansion of these
diagrams is described in \cite{DST}, and there are no remaining
``holes'' there. It is interesting to note that all these algorithms
can be applied to two-loop three-point functions as well.

The general problem of describing the threshold behaviour is still waiting
for a solution. Here we have considered only a part of this problem,
corresponding to the zero-threshold cases. In principle, the
general theory of asymptotic expansions can be applied in some regions
between the thresholds, whilst the description of the behaviour near the
non-zero thresholds requires other approaches.

\vspace{6mm}

A.~D. would like to thank the Instituut-Lorentz, University of Leiden,
and J.B.~T. the Department of Physics, University of Bergen,
for hospitality during visits when parts of this work were done.
We are grateful to A.~Czarnecki and J.~Fujimoto for their
help in comparisons with results based on other numerical
approaches. A.~D.'s research was supported by the Research
Council of Norway.

\vspace{3mm}

% ---- \input{app_a}
\begin{center}
{\bf Appendix A. Two-loop massive vacuum diagrams}
\end{center}

To evaluate two-loop vacuum massive diagrams
occurring in the separate terms (\ref{c1-1})--(\ref{c4-3}) contributing to
the zero-threshold
expansion, we mainly use the
algorithms and formulae presented in \cite{DT,DST}.

The integrals we are interested in are defined as
\be
\label{defI}
I(\nu_1, \nu_2, \nu_3; m_1, m_2, m_3)
\equiv
\int \int \frac{\mbox{d}^n p \; \mbox{d}^n q}
               {[p^2 - m_1^2]^{\nu_1} [q^2 - m_2^2]^{\nu_2}
                   [(p-q)^2 - m_3^2]^{\nu_3}} .
\ee
The general case of this integral with arbitrary $n$ and $\nu$'s was
considered in \cite{DT}. By use of the technique \cite{BD}, a result in
terms of hypergeometric functions of two variables was obtained.
For the case when all $\nu$'s are equal to one and $n$ is arbitrary,
this result was reduced to hypergeometric functions $_2 F_1$ of one variable
(see also in \cite{sch}). Expanding in $\ep$, we get standard
results in terms of dilogarithms or Clausen's function given by
eqs.~(4.9), (4.10) and (4.15) of \cite{DT} (see also in
\cite{Bij,Jones}).

In this paper, however, we always have at least one of the masses
in (\ref{defI}) equal to zero. Generally, this simplifies the
calculations but requires some changes in the algorithms as
compared with the general mass case \cite{DT}.

If one of the masses (e.g. $m_3$) is zero, the result for the
integral (\ref{defI}) can be expressed in terms of a Gauss
hypergeometric function,
\bea
\label{2F1}
I(\nu_1, \nu_2, \nu_3; m_1, m_2, 0)
= \pi^n i^{2-2n} (-m_2^2)^{n-\nu_1 -\nu_2 - \nu_3}
\hspace{40mm}
\nonumber \\[2mm]
\times \frac{\Gamma\left(\frac{n}{2}-\nu_3\right)
             \Gamma\left(\nu_1 + \nu_3 - \frac{n}{2}\right)
             \Gamma\left(\nu_2 + \nu_3 - \frac{n}{2}\right)
             \Gamma\left(\nu_1 + \nu_2 + \nu_3 - n\right)}
            {\Gamma\left(\nu_1\right)
             \Gamma\left(\nu_2\right)
             \Gamma\left(\frac{n}{2}\right)
             \Gamma\left(\nu_1 + \nu_2 + 2\nu_3 - n\right)}
\hspace{10mm}
\nonumber \\[2mm]
\times
\; _2 F_1 \left.
          \left(  \begin{array}{c}{\nu_1 + \nu_3 - \frac{n}{2}, \;
                                   \nu_1 + \nu_2 + \nu_3 - n} \\
                                  {\nu_1 + \nu_2 + 2\nu_3 - n}
                  \end{array}
          \right| 1 - \frac{m_1^2}{m_2^2}  \right) ,
\eea
where it is understood that $i^{-2n} (-m^2)^n = (m^2)^n$.
This result can easily be obtained using the formulae of \cite{BD,DT}.
The symmetry $(m_1, \nu_1) \leftrightarrow (m_2, \nu_2)$ can be
seen by use of the well-known transformation
\be
_2 F_1 \left.
          \left(  \begin{array}{c}{a, \; b} \\ { c }
                  \end{array}
          \right| z \right)
= (1-z)^{-b} \;
 _2 F_1 \left.
          \left(  \begin{array}{c}{c-a, \; b} \\ { c }
                  \end{array}
          \right| \frac{z}{z-1} \right).
\ee
For the cases when $m_1=m_2$ or $m_2=0$ the formula (\ref{2F1})
gives the correct answers:
\bea
\label{mm0}
I(\nu_1, \nu_2, \nu_3; m, m, 0)
= \pi^n i^{2-2n} (-m^2)^{n-\nu_1 -\nu_2 - \nu_3}
\hspace{50mm}
\nonumber \\[2mm]
\times \frac{\Gamma\left(\frac{n}{2}-\nu_3\right)
             \Gamma\left(\nu_1 + \nu_3 - \frac{n}{2}\right)
             \Gamma\left(\nu_2 + \nu_3 - \frac{n}{2}\right)
             \Gamma\left(\nu_1 + \nu_2 + \nu_3 - n\right)}
            {\Gamma\left(\nu_1\right)
             \Gamma\left(\nu_2\right)
             \Gamma\left(\frac{n}{2}\right)
             \Gamma\left(\nu_1 + \nu_2 + 2\nu_3 - n\right)} ;
\hspace{10mm}
\eea
\bea
\label{m00}
I(\nu_1, \nu_2, \nu_3; m, 0, 0)
= \pi^n i^{2-2n} (-m^2)^{n-\nu_1 -\nu_2 - \nu_3}
\hspace{50mm}
\nonumber \\[2mm]
\times \frac{\Gamma\left(\frac{n}{2}-\nu_2\right)
             \Gamma\left(\frac{n}{2}-\nu_3\right)
             \Gamma\left(\nu_2 + \nu_3 - \frac{n}{2}\right)
             \Gamma\left(\nu_1 + \nu_2 + \nu_3 - n\right)}
            {\Gamma\left(\nu_1\right)
             \Gamma\left(\nu_2\right)
             \Gamma\left(\nu_3\right)
             \Gamma\left(\frac{n}{2}\right)} .
\hspace{10mm}
\eea
These results are well-known, and we present them here for
completeness only.

For the case when all $\nu$'s are equal to one and $n \to 4 \; (\ep \to 0)$,
the result for (\ref{2F1}) can be written as
\bea
\label{I(1,1,1)}
I(1, 1, 1; m_1, m_2, 0)
= \pi^{4-2\ep} \; \frac{\Gamma^{2}(1+\ep)}{2 (1-\ep)(1-2\ep)}
\hspace{65mm}
\nonumber \\[2mm]
\times
\left\{ - \frac{1}{\ep^2} (m_1^2 + m_2^2)
               + \frac{2}{\ep} (m_1^2 \ln(m_1^2) + m_2^2 \ln(m_2^2))
\right.
\hspace{50mm}
\nonumber \\[2mm]
\left.
- 2 (m_1^2 \ln^2(m_1^2) + m_2^2 \ln^2(m_2^2))
+ \frac{1}{2} (m_1^2 + m_2^2) \ln^2 \frac{m_2^2}{m_1^2}
+ (m_1^2 - m_2^2) {\cal{H}}(m_1^2,m_2^2)
\frac{}{} \right\},
\eea
where ${\cal{H}}$ is a dimensionless antisymmetric function defined as
\be
\label{defH-2}
{\cal{H}}(m_1^2, m_2^2) = 2\Li{2}{1- \frac{m_1^2}{m_2^2}}
               + \frac{1}{2} \ln^2\left(\frac{m_1^2}{m_2^2}\right).
\ee
It is connected with the function $F$ used in \cite{DST} (see Appendix B,
eqs.~(B.3)--(B.7) of \cite{DST}) via
\be
{\cal{H}}(m_1^2,m_2^2)
=
- \begin{array}{c}{\frac{}{}}\\{\mbox{lim}}\\{{}_{m_3 \to 0}} \end{array}
\left\{ \frac{F(m_1^2, m_2^2, m_3^2)}{m_1^2 - m_2^2}
        - \frac{1}{2} \ln\frac{m_1^2}{m_2^2}
\; \left( \ln\frac{m_1^2}{m_3^2} + \ln\frac{m_2^2}{m_3^2} \right)
\right\} .
\ee
We cannot use the original function $F$ here because it has a logarithmic
singularity as $m_3 \to 0$.

We need to calculate integrals (\ref{defI}) with arbitrary (integer)
powers of denominators. For the cases when we have one mass parameter
only, it is convenient to use the simple explicit formulae
(\ref{mm0}), (\ref{m00}).
If we have different masses, however, it is better to use recurrence
relations obtained by using the integration-by-parts technique
\cite{ibp} (in a way analogous to \cite{Davyd-JPA}).
For the case of the integrals (\ref{defI}),
these relations were considered in \cite{DT}.
The solution to these relations presented there (see eq.~(5.3)
of \cite{DT}) contains masses in the denominator. This can be avoided
by employing the connection (5.4) of the same paper.
Note that for
$m_3=0$ the determinant of the system of equations corresponding to
the recurrence relations becomes
$\Delta(m_1^2, m_2^2, 0) = - (m_1^2 - m_2^2)^2$.

Of course, application of these recurrence relations is equivalent
to using connections between contiguous hypergeometric functions
$_2 F_1$ (see, e.g., in \cite{Erdelyi}) corresponding to
eq.~(\ref{2F1}). Furthermore, in cases
%when the values of the masses are close together
when we have close values of the masses, it may be better to use the
explicit result (\ref{2F1}) to expand in the mass difference
up to the order corresponding to the required accuracy.

\vspace{3mm}

% ---- \input{app_b}
\begin{center}
{\bf Appendix B. Two-loop massless integrals with numerators}
\end{center}

Here we shall consider massless integrals
\be
\label{defJ0}
J^{(0)}(\nu_1, \nu_2, \nu_3, \nu_4, \nu_5)
\equiv J\left( \{\nu_i\} ; 0,0,0,0,0 ; k \right),
\ee
where $J$ corresponds to the original integral (\ref{defJ}).
The main properties of these integrals are well-known,
and many of them can be obtained by use of the integration-by-parts
technique \cite{ibp} (we had collected some of them in Appendix A
of \cite{DST}).
In particular, when any of $\nu$'s is equal to zero, the corresponding
integral can be calculated in terms of gamma functions,
for any value of the space-time dimension $n$.
If all $\nu$'s are equal to one and $n=4$, the result
is well-known ($-6 \zeta(3) \pi^4/k^2$).
For higher powers of propagators, integration by parts \cite{ibp}
can be used, which was implemented in programs like \cite{mincer}.

In this appendix, we would like to discuss a ``non-standard'' way
of calculating the integrals in cases when some of the $\nu$'s
are negative. So, in fact, we get the corresponding denominators in the
numerator. This approach is based on the formula presented in
ref.~\cite{Davyd'91} and was used in this paper. The general
case of this formula can be applied to any one-loop integral with tensor
structure in the numerator. As a result, one gets a sum
of tensor structures multiplied by scalar
integrals in other dimensions. We shall need only a simple corollary
of this general formula below.

In fact, we have two different cases to consider: (i) $\nu_3 < 0$
and (ii) any other $\nu$ (for example, $\nu_5$) is negative.
Let us consider the first possibility, the second one can be
treated analogously.
If $\nu_3 < 0$, this corresponds to getting
\be
\left[ (p-q)^2 \right]^{|\nu_3 |}
\equiv \left[ p^2 + q^2 - 2(pq) \right]^{|\nu_3 |}
\ee
in the numerator. After expanding and cancelling $p^2$ and $q^2$
against the corresponding denominators, we get integrals
of the form
\be
\int \int \mbox{d}^n p \; \mbox{d}^n q \;
\frac{(pq)^M}{\left[ p^2 \right]^{\nu_1}
              \left[ q^2 \right]^{\nu_2}
              \left[ (k-p)^2 \right]^{\nu_4}
              \left[ (k-q)^2 \right]^{\nu_5}} .
\ee

Now, let us write $(pq)^M$ as
$p^{\mu_1} \ldots p^{\mu_M} \; q_{\mu_1} \ldots q_{\mu_M}$ and
consider one of the subloops (an analogy with
\cite{Weigl} can be noticed). According to \cite{Davyd'91}, it can
be written as
\footnote{There was a misprint in the corresponding formula
(17) of \cite{Davyd'91}: the last gamma function on the first line
of eq.~(17) should be $\Gamma(n-\nu_1-\nu_2+M)$.
The general formula (11) is correct.}
\bea
\label{tensor}
\int \mbox{d}^n q \;
\frac{q_{\mu_1} \ldots q_{\mu_M}}
     {\left[ q^2 \right]^{\nu_2} \; \left[ (k-q)^2 \right]^{\nu_5}}
= \pi^{n/2} \mbox{i}^{1-n} (k^2)^{n/2 -\nu_2 - \nu_5}
  \left[ \Gamma(\nu_2) \Gamma(\nu_5) \Gamma(n -\nu_2 - \nu_5 + M)
       \right]^{-1}
\nonumber \\[2mm]
\times
\sum\limits_{\begin{array}{c}
               {{}_{\lambda, \; \kappa}}\\{{}_{2\lambda + \kappa = M}}
             \end{array}}^{}
\left( \frac{k^2}{2} \right)^{\lambda}
\left\{ [g]^{\lambda} [k]^{\kappa} \right\}_{\mu_1 ... \mu_M}
\Gamma\left(\!\frac{n}{2}\! -\! \nu_2 \!+\! \lambda\! + \!\kappa \!\right)
\Gamma\left(\!\frac{n}{2}\! - \!\nu_5 \!+\! \lambda\!\right)
\Gamma\left(\!\nu_2 \!+ \!\nu_5 \!- \!\frac{n}{2}\! -\! \lambda\!\right),
\eea
where $\left\{ [g]^{\lambda} [k]^{\kappa} \right\}_{\mu_1 \ldots \mu_M}$
denotes a symmetric tensor structure each term of which
is constructed from $\lambda$ metric tensors $g_{\mu_i \mu_j}$
and $\kappa$ vectors $k_{\mu_i}$, the coefficient of each
term being equal to one. It is easy to show that such a tensor
structure contains
\be
\label{comb}
\frac{(2\lambda + \kappa)!}{2^{\lambda}\;{\lambda}! \; {\kappa}!}
\ee
terms.
Moreover, when contracted with $p^{\mu_1} \ldots p^{\mu_M}$,
each term gives the same result, namely
$(p^2)^{\lambda} \; (kp)^{\kappa}$.
% Moreover, each term gives the same result when we contract it
%with $p^{\mu_1} \ldots p^{\mu_M}$ remaining in the numerator.
%Namely, it gives $(p^2)^{\lambda} \; (kp)^{\kappa}$.

So, now we obtain a sum of the terms of the form
\be
\int \mbox{d}^n p \; \frac{(kp)^{\kappa}}
                          {\left[ p^2 \right]^{\nu_1 - \lambda} \;
                           \left[ (k-p)^2 \right]^{\nu_4}} ,
\ee
with known coefficients (\ref{comb}). Finally, the scalar products
$(kp)$ in the numerator can be represented in terms of the
denominators and $k^2$, and the resulting integrals are trivial.

\vspace{3mm}

% ---- \input{refs}

\end{document}